\documentclass[prl,twocolumn,floatfix]{revtex4}
\usepackage{amssymb}
\usepackage{amsmath}
\usepackage{epsfig}
\usepackage{graphicx}

\begin{document}

\title{ Oscillation bands of condensates on a ring: \ Beyond the mean field
theory}
\author{C. G. Bao}
\affiliation{Center of Theoretical Nuclear Physics, National Laboratory of Heavy Ion
Collisions, Lanzhou 730000, P. R. China \\
and \\
The State Key Laboratory of Optoelectronic Materials and Technologies,
Zhongshan University, \ Guangzhou, 510275, P.R. China}

\begin{abstract}
Abstract: The Hamiltonian of a $N$-boson system confined on a ring with zero
spin and repulsive interaction is diagonalized. The excitation of a pair of
p-wave-particles rotating reversely appears to be a basic mode. The
fluctuation of many of these excited pairs provides a mechanism of
oscillation, the states can be thereby classified into oscillation bands. \
The particle correlation is studied intuitively via the two-body densities.
Bose-clustering originating from the symmetrization of wave functions is
found, which leads to the appearance of 1-, 2-, and 3-cluster structures. \
The motion is divided into being collective and relative, this leads to the
establishment of a relation between the very high vortex states and the
low-lying states.
\end{abstract}

\maketitle

After the experimental realization of the Bose-Einstein condensation$^{1}$,
various condensates confined under different circumstances have been
extensively studied theoretically and experimentally. \ Mostly, the
condensates are considered to be confined in a harmonic trap. \ Condensates
trapped by periodic potential have also been studied due to the appearance
of optical lattices.$^{2}$ \ It is believed that the appearance of
condensates confined in particular geometries is possible. Experimentally,
the particle interactions can now be tuned from very weak to very strong,$%
^{3-8}$ it implies that the particle correlation may become important. \
Theoretically, to respond, going beyond the mean field Gross-Pitaevskii (GP)
theory is desirable, and the condensates confined in particular geometries
are also deserved to be considered.

Along this line, in addition to the ground state, the yrast states have been
studied both analytically and numerically.$^{9-17}$ The condensation on a
ring has also been studied recently.$^{12}$ \ The present paper is also
dedicated to the $N-$boson systems confined on a ring with weak
interaction,\ its scope is broader and covers the whole low-lying spectra. \
A similar system has been investigated analytically by Lieb and Liniger$%
^{16,17}$. However, the emphasis of their papers is different from the
present one, which is placed on analyzing the structures of the excited
states to find out their distinctions and similarities, and to find out the
modes of excitation. \ Based on the analysis, an effort is made to classify
the excited states. Traditionally, the particle correlation and its effect
on the geometry of $N-$boson systems is a topic scarcely studied if $N$ is
large. \ In this paper, the correlation is studied intuitively so as the
geometric features inherent in the excited states can be understood.
Traditionally, a separation between the collective and internal motions is
seldom to be considered if $N$ is large. \ In this paper such a separation
is made and leads to the establishment of a relation between the vortex
states and the low-lying states.\ \

It is assumed that the $N$ identical bosons confined on a ring have mass $m,$
spin zero, and square-barrier interaction. The ring has a radius $R$, $N$ \
is given at 100, 20 and 10000. \ \ Let $G=\hbar ^{2}/(2mR^{2})$ be the unit
of energy. The Hamiltonian then reads
\begin{equation}
\begin{array}{lll}
H=-\sum_{i}\frac{\partial ^{2}}{\partial \theta _{i}^{2}}+\sum_{i<j}V_{ij} &
&
\end{array}%
\end{equation}%
where $\theta _{i}$\ is the azimuthal angle of the i-th boson. $V_{ij}=V_{o}$
if $|\theta _{j}-\theta _{i}|\leq \theta _{range}$ , or $=0$ otherwise. \
Let $\phi _{k}=e^{ik\theta }/\sqrt{2\pi }$ be a single particle state, $%
-k_{\max }\leq k\leq k_{\max }$\ is assumed. \ The $N-$body normalized basis
functions in Fock-representation are $|\alpha \rangle \equiv |n_{-k_{\max
}},\cdot \cdot \cdot \ n_{k_{\max }}\rangle ,$ where $n_{j}$\ is the number
of bosons in $\phi _{j}$, $\sum_{j}n_{j}=N,$ and $\sum_{j}n_{j}j=L$, the
total angular momentum. Then, $H$ is diagonalized in the space spanned by $%
|\alpha \rangle ,$\ the low-lying spectrum together with the
eigen-wave-functions, each is a linear combination of $|\alpha \rangle $,
are thereby obtained. \ Let $K_{\alpha }=\sum_{j}n_{j}j^{2}$\ be the total
kinetic energy of an $|\alpha \rangle $ state. Evidently, those $|\alpha
\rangle $\ with a large $K_{\alpha }$\ are negligible for low-lying states.
Therefore, one more constraint $K_{\alpha }\leq K_{\max }$\ is further added
to control the number of $|\alpha \rangle $. In this procedure, the crucial
point is the calculation of the matrix elements of $H$. This can be realized
by using the fractional parentage coefficients$^{18}$ (refer to eq.(6)
below). Numerical results are reported as follows.

This paper concerns only the cases with weak interaction. Firstly, let $%
V_{o}=1,\theta _{range}=0.025$, and $N=100$. This is corresponding to $%
\gamma =0.00157$ , where $\gamma $\ is introduced by Lieb and Liniger to
measure the strength of interaction,$^{16,17}$ this is shown later. When $%
k_{\max }$ and $K_{\max }$ are given at a number of values, the associated
eigen-energies $E_{j}$ of the first, fifteenth, and sixteenth $L=0$
eigen-states are listed in Table I. \ When $(k_{\max },K_{\max })$ is
changed from $(3,50)$ to (5,60), the total number of $|\alpha \rangle $\ is
changed from 2167 to 8890. Table I demonstrates that the great increase of
basis functions does not lead to a remarkable decrease of eigen-energies.
Thus the convergency is qualitatively satisfying even for the higher states.
\begin{table}[tbph]
\caption{Eigen-energies $E_{j}$ (the unit is $G$) of the $L=0$ states. The
first row is $(k_{\max },K_{\max })$, the first column is the serial number
of states $j$. $V_{o}=1,\protect\theta _{range}=0.025,$ and $N=100$ are
given.}%
\begin{ruledtabular}
\begin{tabular}{cccccc}
   & (3,50) & (4,50) & (4,60) & (5,60) \\ \hline
 1 & 39.109 & 39.090 & 39.090 & 39.078 \\
15 & 53.645 & 53.616 & 53.616 & 53.613 \\
16 & 54.822 & 54.800 & 54.800 & 54.790
\end{tabular}
\end{ruledtabular}
\end{table}

In the following the choice $k_{\max }=4$ and $K_{\max }=50$ are adopted,
this limitation leads to a 3254-dimensional space. \ Thereby the resultant
data have at least three effective figures, this is sufficient for our
qualitative purpose.
\begin{figure}[htbp]
\centering
\includegraphics[totalheight=1.4in,trim=30 10 5 10]{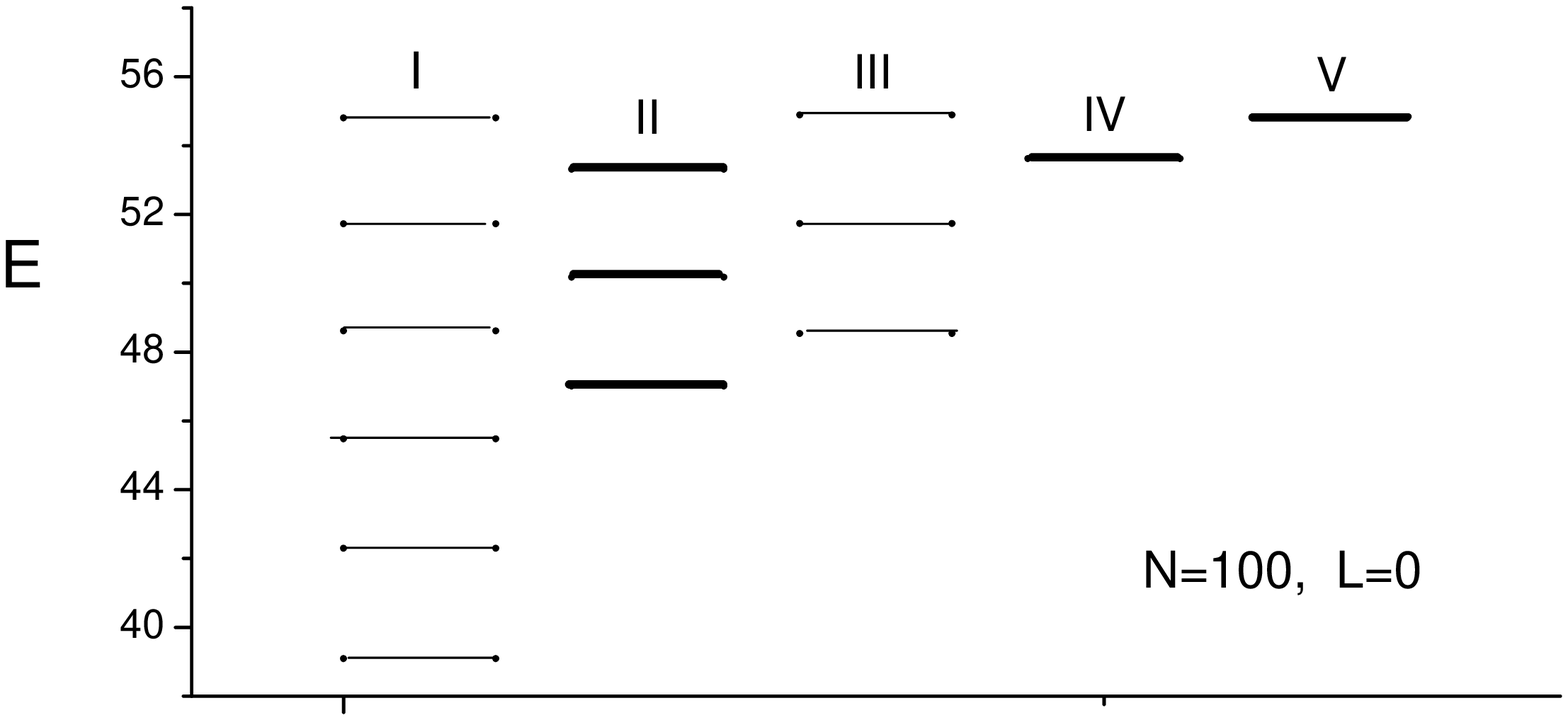}
\caption{The spectrum of $L=0$ states, the unit of energy is $G=\hbar
^{2}/(2mR^{2})$. $N=100$, $V_{o}=1,$ and $\protect\theta _{range}=0.025$ are
assumed, they are the same for Fig.1 to Fig.5. The levels in a column
constitute an oscillation band, the levels in bold line are doubly
degenerate.}
\end{figure}

The low-lying spectrum is given in Fig.1, where the lowest fourteen levels
are included. Twelve of them can be ascribed into three bands, in each band
the levels are distributed equidistantly, this is a strong signal of
harmonic-like oscillations. From now on the labels $\Psi _{Z,i}^{(L)}$\ and $%
E^{(L,Z,i)}$\ are used to denote the wave function and energy of the i-th
state of the Z-th band (Z=I, II, III,$\cdot \cdot \cdot $).

It turns out that the excitation of a pair of particles both in p-wave but
rotating reversely, namely, one particle in $\phi _{1}$ while the other one
in $\phi _{-1},$ is a basic mode, the pair is called a basic pair in the
follows. A number of such basic pairs might be excited. \ When $2j$
particles are in basic pairs while the remaining $N-2j$ particles are in $%
\phi _{0}$, the associated $|\alpha \rangle $ is written as $|P^{(j)}\rangle
$. For all the states of the $I-$band, we found $\Psi _{I,i}^{(0)}$ is
mainly a linear combination of $|P^{(j)}\rangle $\ together with a small
component denoted by $\Delta _{I,i}$, i.e.,
\begin{equation}
\Psi _{I,i}^{(0)}=\sum_{j}C_{j}^{(0,I,i)}\ |P^{(j)}\rangle +\Delta _{I,i}
\end{equation}%
where $\Delta _{I,i}$ is very small as shown in Table II, while the
coefficients $C_{j}^{(L,Z,i)}$ arise from the diagonalization. Thus the
basic structure of the $I-$band is just a fluctuation of many of the basic
pairs.
\begin{table}[tbph]
\caption{The weights of $\Delta _{Z,i}$ of the bands with $L=0$}%
\begin{ruledtabular}
\begin{tabular}{cccc}
$i$ & $I-$band & $II-$band & $III-$band \\ \hline
1 & 0.009 & 0.017 & 0.040 \\
2 & 0.012 & 0.030 & 0.056 \\
3 & 0.021 & 0.061 & 0.088 \\
4 & 0.035 & 0.028 &  \\
5 & 0.055 & 0.035 &  \\
6 & 0.079 & 0.106 &
\end{tabular}
\end{ruledtabular}
\end{table}

For lower states, $C_{j}^{(0,Z,i)}$ would be very small if $j$ is larger,
e.g., for the ground state, $C_{0}^{(0,I,1)}=0.968$ and$\ C_{j\ \geq
2}^{(0,I,1)}\approx 0,$ it implies that the excitation of many pairs is not
probable. It also implies that the ground state wave function obtained via
mean-field theory might be a good approximation. \ However, for higher
states, many pairs would be excited. \ E.g., for the third state of the $I-$%
band, $C_{j}^{(0,I,3)}=0.052$,$\ 0.406$,$\ 0.681,$ $-0.528,$ and $0.245$\
when $j$ is from 0 to 4, it implies a stronger fluctuation.

When a $|\alpha \rangle $ has not only $2j$\ particles in the basic pairs,
but also$\ m$ particles in $\phi _{k}$, while the remaining particles in $%
\phi _{0}$, then it is denoted as $|(k)^{m}P^{(j)}\rangle $ (where $k=\pm 1$
are allowed) \ Similarly, we can define $%
|(k_{1})^{m_{1}}(k_{2})^{m_{2}}P^{(j)}\rangle $, and so on. \ For all the
states of the $II-$band, we found
\begin{equation}
\begin{array}{lll}
\Psi _{II,i}^{(0)} & = & \sum_{j}C_{j}^{(0,II,i)}\frac{1}{\sqrt{2}}[\
|(2)^{1}(-1)^{2}P^{(j)}\rangle \\
&  & \pm |(-2)^{1}(1)^{2}P^{(j)}\rangle ]+\Delta _{II,i}%
\end{array}%
\end{equation}%
where both the $+$ and $-$ signs lead to the same energy, thus the level is
two-fold degenerate. Again, all the $\Delta _{II,i}$ are very small as shown
in Table II, \ thus the fluctuation of basic pairs is again the basic
structure. However, the $II-$band is characterized by having the additional
3-particle-excitation (one in d-wave and two in p-wave).

For all the states of the third band, we found
\begin{equation}
\begin{array}{lll}
\Psi _{III,i}^{(0)}=\sum_{j}C_{j}^{(0,III,i)}\
|(2)^{1}(-2)^{1}P^{(j)}\rangle +\Delta _{III,i} &  &
\end{array}%
\end{equation}%
Thus, the $III-$band contains, in addition to the fluctuation of basic
pairs, a more energetic pair with each particle in d-wave. It was found that
the spacing $E^{(0,Z,i+1)}-E^{(0,Z,i)}$ inside all the bands are nearly the
same, they are\ $\sim $3.15. \ This arises because they have the same
mechanism of oscillation, namely, the fluctuation of basic pairs.

When the energy goes higher, more oscillation bands can be found. \ The two
extra levels in Fig.1 at the right are the band-heads of higher bands.

Incidentally, the band-heads of the above three bands are dominated by $%
|P^{(0)}\rangle ,$ $|(2)^{1}(-1)^{2}P^{(0)}\rangle \pm
|(-2)^{1}(1)^{2}P^{(0)}\rangle $ and $|(2)^{1}(-2)^{1}P^{(0)}\rangle $,
respectively, and their kinetic energies $K_{\alpha }=0,6,$ and 8. \ Among
all the basis functions with $L=0$\ and without basic pairs, these three are
the lowest three. \ This explains why the band-heads are dominated by them.
Once a band-head is fixed, the corresponding oscillation band would grow up
via the fluctuation of basic pairs.

The particle correlations can be seen intuitively by observing the two-body
densities
\begin{equation}
\begin{array}{lll}
\rho _{2}(\theta _{1},\theta _{2})=\int d\theta _{3}\cdot \cdot \cdot
d\theta _{N}\ \Psi _{Z,i}^{(L)\ast }\Psi _{Z,i}^{(L)} &  &
\end{array}%
\end{equation}

Similar to the calculation of the matrix elements of interaction, the above
integration can be performed in coordinate space by extracting the particles
1 and 2 from $|\alpha \rangle $\ by using the fractional parentage
coefficients$^{18}$, namely,
\begin{equation}
\begin{array}{rll}
|\alpha \rangle =\sum_{k}\sqrt{n_{k}(n_{k}-1)/N(N-1)}\phi _{k}(1)\phi
_{k}(2)|\alpha _{k}\rangle &  &  \\
+\underset{(k_{a}\neq k_{b})}{\sum_{k_{a},k_{b}}}\sqrt{%
n_{k_{a}}n_{k_{b}}/N(N-1)}\phi _{k_{a}}(1)\phi _{k_{b}}(2)|\alpha
_{k_{a}k_{b}}\rangle &  &
\end{array}%
\end{equation}%
where $|\alpha _{k}\rangle $ is different from $|\alpha \rangle $ by
replacing $n_{k}$ with $n_{k}-2$, $\ |\alpha _{k_{a}k_{b}}\rangle $ is
different from $|\alpha \rangle $ by replacing $n_{k_{a}}$ and $n_{k_{b}}$
with $n_{k_{a}}-1$ and $n_{k_{b}}-1,$ respectively.
\begin{figure}[tbph]
\centering
\includegraphics[totalheight=2.2in,trim=30 10 5 10]{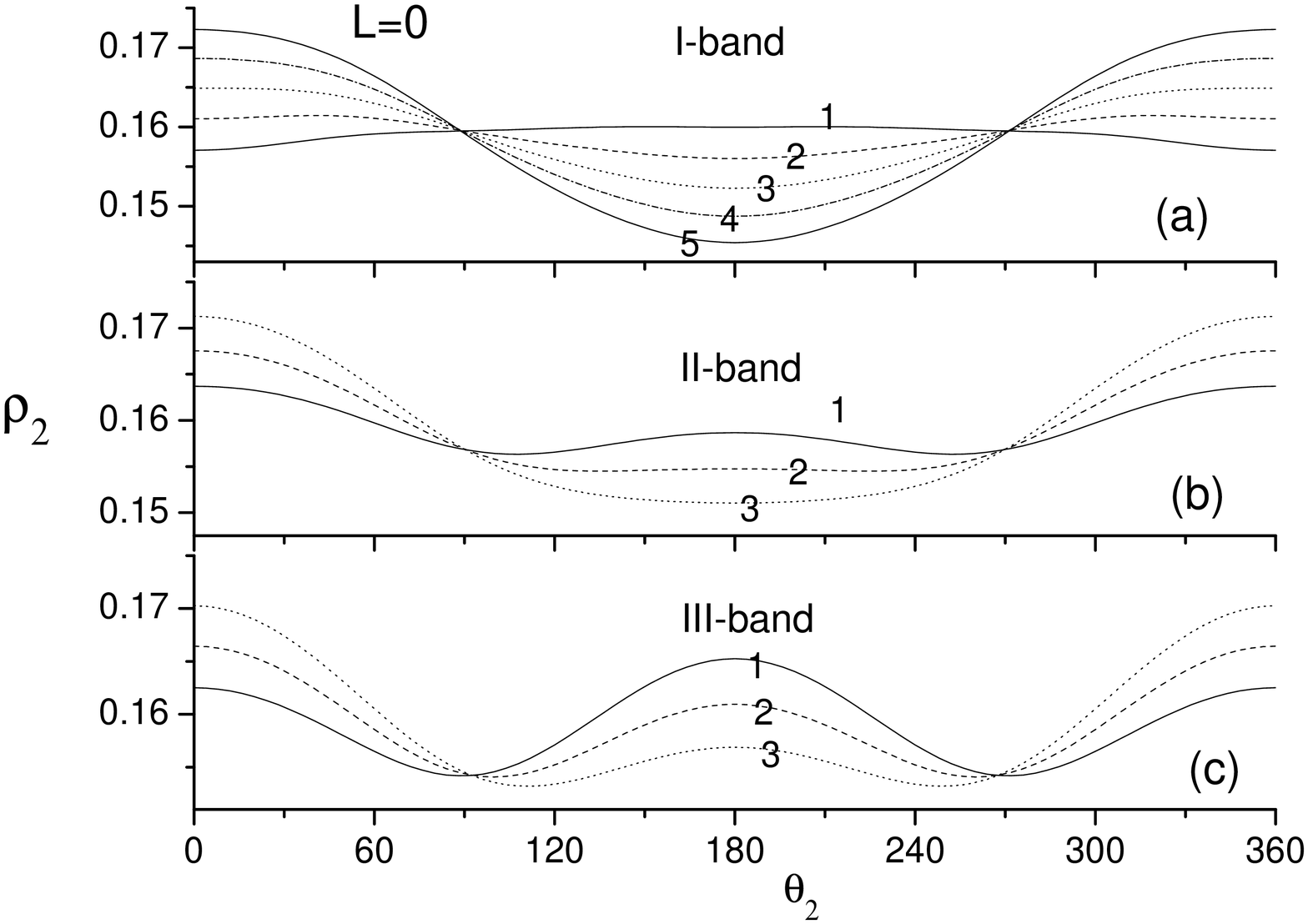}
\caption{$\protect\rho _{2}$\ as functions of $\protect\theta _{2}$\ for the
$I\ ($ $a),\ II\ ($ $b),\ $and $III\ (c)$ bands of $L=0$ states, $\protect%
\theta _{1}=0$ is given. \ The labels $i$ of the states $\Psi _{Z,i}^{(0)}$
are marked by the curves.}
\end{figure}

$\rho _{2}$ gives the spatial correlation between any pair of particles as
shown in Fig.2. For the ground state $\Psi _{I,1}^{(0)}$, $\rho _{2}$ is
flat implying that the correlation is weak. However, it\ is a little larger
when the two particles are opposite to each other ($\theta _{1}=0$ and $%
\theta _{2}=\pi $). It implies the existence of a weak correlation which is
entirely ignored by the mean field theory. \ Thus, even the interaction
adopted is weak and even for the ground state, there is still a small
revision to the mean field theory.\ For higher states of the $I-$band, the
fluctuation of basic pairs becomes stronger. \ Due to the fluctuation, the
particles tend to be close to each other to form a single cluster. This
tendency is clearly shown in Fig.2a.

For the first state of the $III-band,$ $\Psi _{III,1}^{(0)}$ has two peaks
in $\rho _{2}$ implying a 2-cluster structure. It arises from the two d-wave
paticles inherent in the band. \ The feature of $\Psi _{II,1}^{(0)}$ is
lying between $\Psi _{I,1}^{(0)}$ and $\Psi _{III,1}^{(0)}$. \ For all
higher states of every band, due to the strong fluctuation of basic pairs,
all the particles tend to be close to each other as shown in 2b and 2c.

To understand the physics why the particles tend to be close to each other,
let us study the most important basis state $|P^{(j)}\rangle $. \ By
inserting $|P^{(j)}\rangle $ into eq.(5) to replace $\Psi _{Z,i}^{(L)}$ and
by using (6), $\rho _{2}$\ reads
\begin{equation}
\begin{array}{rll}
\rho _{2}(\theta _{1},\theta _{2})=\frac{1}{(2\pi )^{2}N(N-1)}
[N(N-1)-j(4N-6j) &  &  \\
+4j(N-2j)(1+\cos (\theta _{1}-\theta _{2}))+4j^{2}\cos ^{2}(\theta
_{1}-\theta _{2})] &  &
\end{array}%
\end{equation}
Where there are four terms at the right, the non-uniformity arises from the
third and fourth terms. The third term causes the particles to be close to
each other to form a single cluster, while the fourth term causes the
two-cluster clustering. When $j$ is small, the fourth term can be neglected,
and the particles tend to form a single cluster. However, when $j\approx N/2$
, the third term can be neglected, and the particles tend to form two
clusters. It is noted that, if the symmetrization were dropped, the density
contributed by $|P^{(j)}\rangle $ would be uniform. The appearance of the
clustering originates from the symmetrization of the bosonic wave functions,
therefore it can be called as bose-clustering.

For $L=1$ states,\ the lowest energy $E^{(1,I,1)}$ is higher than $%
E^{(0,I,1)}$ by 1.606, but lower than $E^{(0,I,2)}$. \ Thus $\Psi
_{I,1}^{(1)}$ is the true \ first excited state of the system. A number of
oscillation bands exist as well, the wave functions of the lowest six bands
are found as
\begin{equation}
\begin{array}{lll}
\Psi _{I,i}^{(1)}=\sum_{j}C_{j}^{(1,I,i)}\ |(1)^{1}P^{(j)}\rangle +\Delta
_{I,i} &  &  \\
\Psi _{II,i}^{(1)}=\sum_{j}C_{j}^{(1,II,i)}\ |(2)^{1}(-1)^{1}P^{(j)}\rangle
+\Delta _{II,i} &  &  \\
\Psi _{III,i}^{(1)}=\sum_{j}C_{j}^{(1,III,i)}\
|(-2)^{1}(1)^{3}P^{(j)}\rangle +\Delta _{III,i} &  &  \\
\Psi _{IV,i}^{(1)}=\sum_{j}C_{j}^{(1,IV,i)}\
|(2)^{1}(-2)^{1}(1)^{1}P^{(j)}\rangle +\Delta _{IV,i} &  &  \\
\Psi _{V,i}^{(1)}=\sum_{j}C_{j}^{(1,V,i)}\ |(3)^{1}(-1)^{2}P^{(j)}\rangle
+\Delta _{V,i} &  &  \\
\Psi _{VI,i}^{(1)}=\sum_{j}C_{j}^{(1,VI,i)}\ |(3)^{1}(-2)^{1}P^{(j)}\rangle
+\Delta _{VI,i} &  &
\end{array}%
\end{equation}%
Where the weights of all the $\Delta _{Z,i}\leq $ 0.1 if $i\leq 4$. \ Thus,
just as the above $L=0$ case, all the bands have the common fluctuation of
basic pairs, but each band has a specific additional few-particle
excitation. The energies of the band-heads from $I$ to $VI$ are 40.70,
45.42, 48.58, 50.14, 52.04, and 53.58 respectively. \ Furthermore, the
spacing $\sim $3.15 found above is found again for all these bands due to
having the same mechanism of oscillation. \ The $I-$band is similar to the
above $I-$band with $L=0$\ but having an additional single p-wave
excitation, the $\rho _{2}$ of them are one-one similar. Similarly, the $%
\rho _{2}$ of the $IV-$band is one-one similar to those of the above $III-$
band with $L=0$. The $\rho _{2}$ of the $II$ and $III-$bands are both
similar to those of the above $II-$band with $L=0$. \ However, the $V$ and $%
VI-$bands are special due to containing the f-wave excitation, the $\rho
_{2} $ of their band-heads exhibit a 3-cluster structure as shown in Fig.3.
\ \ When the energy goes even higher, more higher oscillation bands will
appear. \ For the above six $L=1$ bands, their band-heads are dominated by
the $|\alpha \rangle $ with $K_{\alpha }=1,5,7,9,11,$ and 13. Obviously, a
higher $K_{\alpha }$ leads to a higher band.
\begin{figure}[tbph]
\centering
\includegraphics[totalheight=1.5in,trim=30 10 5 10]{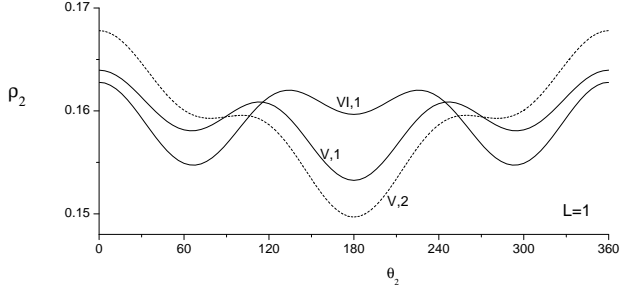}
\caption{$\protect\rho _{2}$ for selected $L=1$ states. \ $\protect\theta %
_{1}=0$, the $(Z,i)$ labels are marked by the curves.}
\end{figure}

In general, all the low-lying states can be classified into oscillation
bands. \ For all the lower bands disregarding $L$, it was found that each
band-head is dominated by a basis function containing a specific
few-particle excitation but not containing any basic pairs. The energy order
of the bands is determined by the magnitudes of $K_{\alpha }$ associated
with the dominant basis function $|\alpha \rangle $ of the band-heads. Once
a band-head stands, an oscillation band will grow up from the band-head
simply via the fluctuation of basic pairs. For examples, for $L=2$ states,
the dominant $|\alpha \rangle $ of the band-heads of the four lowest
oscillation bands are $|(1)^{2}P^{(0)}\rangle $, $|(2)^{1}P^{(0)}\rangle $, $%
|(-2)^{1}(1)^{4}P^{(0)}\rangle $, and $|(3)^{1}(-1)^{1}P^{(0)}\rangle $ with
$K_{\alpha }=2,4,8,$ and $10,$ respectively.

For $L=3$ states, the dominant $|\alpha \rangle $ of the band-heads of the
three lowest bands are $|(1)^{3}P^{(0)}\rangle $, $|(2)^{1}(1)^{1}P^{(0)}
\rangle $, and $|(3)^{1}P^{(0)}\rangle ,$ with $K_{\alpha }=3,5,$ and $9,$
respectively. \ Since the p-, d-, and f-wave appear successively, these
band-heads exhibit 1-cluster, 2-cluster, and 3-cluster structures,
respectively, as shown in Fig.4.
\begin{figure}[htbp]
\centering
\includegraphics[totalheight=1.5in,trim=30 10 5 10]{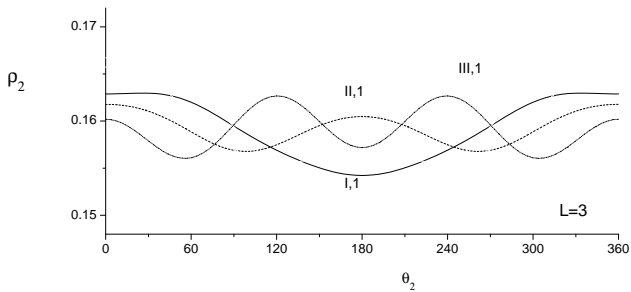}
\caption{$\protect\rho _{2}$ for the band-heads of $L=3$ states, $\protect%
\theta _{1}=0$.}
\end{figure}

Furthermore, a $-L$ state can be derived from the corresponding $L$ state
simply by changing every $k$ to $-k$, i.e., change the components $%
|(k_{1})^{m_{1}}(k_{2})^{m_{2}}P^{(j)}\rangle $ to $%
|(-k_{1})^{m_{1}}(-k_{2})^{m_{2}}P^{(j)}\rangle ,$\ and so on. \ Therefore $%
\Psi _{Z,i}^{(-L)}=(\Psi _{Z,i}^{(L)})^{\ast }$, and $%
E^{(-L,Z,i)}=E^{(L,Z,i)}$.
\begin{figure}[htbp]
\centering
\includegraphics[totalheight=1.5in,trim=30 10 5 10]{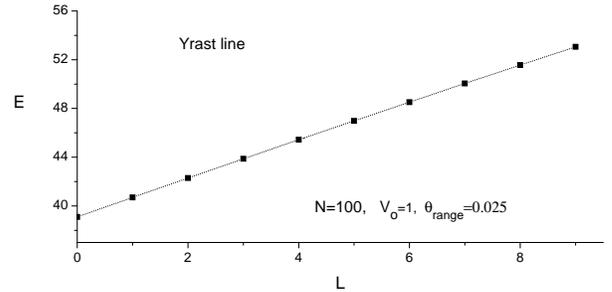}
\caption{Energies of the yrast states with $L=0$ to 10.}
\end{figure}

Let us study the yrast states $\Psi _{I,1}^{(L)}$, each is the lowest one
for a given $L$. The energies of them\ are plotted in Fig.5, their wave
functions are found as
\begin{equation}
\begin{array}{lll}
\Psi _{I,1}^{(L)}=\sum_{j}C_{j}^{(L,I,1)}\ |(1)^{L}P^{(j)}\rangle +\Delta
_{I,1}^{L} &  &
\end{array}%
\end{equation}%
where $\Delta _{I,i}^{L}$ is very small. \ When $L$ is small, the
fluctuation of basic pairs is small, and the yrast states are dominated by
the $j=0$ component $|(1)^{L}P^{(0)}\rangle .$ When $L$ is larger, the
weight of the $|(1)^{L}P^{(0)}\rangle $ component becomes smaller. \ E.g.,
when $L=0,2,4,$ and 10, the weights of $|(1)^{L}P^{(0)}\rangle $ are 0.94,
0.84, 0.75, and 0.54, respectively. \ Evidently, the energy going up
linearly in the yrast line in Fig.5 is mainly due to the linear increase of
the number of p-wave particles.
\begin{figure}[tbph]
\centering
\includegraphics[totalheight=1.4in,trim=30 10 5 10]{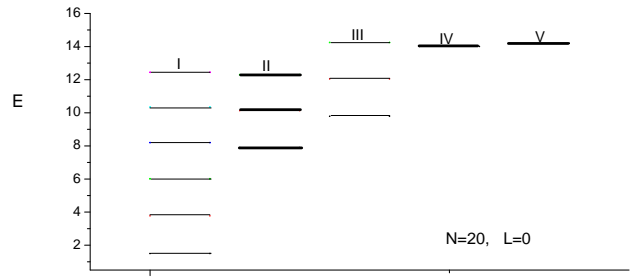}
\caption{The spectrum of the $L=0$ states with $N=20$, $V_{o}=1,$ and $%
\protect\theta _{range}=0.025$. \ Refer to Fig.1}
\end{figure}
\begin{figure}[tbph]
\centering
\includegraphics[totalheight=1.5in,trim=30 10 5 10]{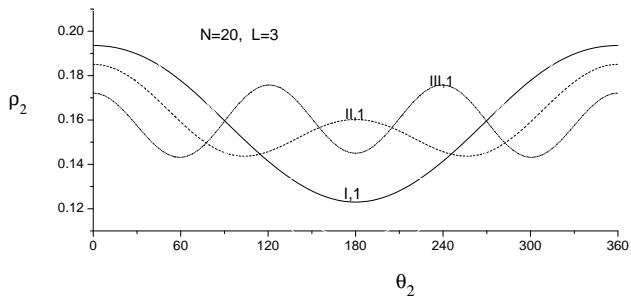}
\caption{The same as Fig.4 but with $N=20$}
\end{figure}

When $N=20$, all the above qualitative features remain unchanged. Examples
are given in Fig.6 and 7 to be compared with Fig.1 and 4. Nonetheless, the
decrease of $N$\ implies that the particles have a less chance to meet each
other, thus the particle correlation is expected to be weaker. \
Quantitatively, it was found that (i) The spacing of adjacent oscillation
levels becomes smaller, it is now $\sim $2.2 to replace the previous 3.15 \
(ii) The fluctuation becomes weaker. \ E.g., the weights of $|P^{(j)}\rangle
$ of the $\Psi _{I,3}^{(0)}$ state are 0.01, 0.95, and 0.03 for $j=1,2,$ and
3, respectively, while these weights would be 0.16, 0.46, and 0.28 if $%
N=100. $ (iii) When $N$ becomes small, the geometric features would become
explicit. \ E.g., for the 3-cluster structure, the difference between the
maximum and minimum of $\rho _{2}$ is $\sim $0.007 in Fig.4, but $\sim $
0.033 in Fig.7. \

The decrease of $V_{o}$ or $\theta _{range}$ was found to cause an effect
similar to the decrease of $N$, the spectra would remain qualitatively
unchanged. \ Quantitatively,\ when $V_{o}$ is changed from 1 to 0.1, the
spacing inside a band is changed from $\sim 3.15$ to $\sim 2.14$, and the
fluctuation becomes much weaker as expected.

In what follows we study the vortex states. For an arbitrary $L_{o}\leq N/2$%
, the spectra of the $\Psi _{Z,i}^{(N-L_{o})}$\ and $\Psi _{Z,i}^{(L_{o})}$
states are found to be identical$^{15}$, except the former shifts upward as
a whole by $N-2L_{o}$, namely,
\begin{equation}
\begin{array}{lll}
E^{(N-L_{o},Z,i)}=E^{(L_{o},Z,i)}+N-2L_{o} &  &
\end{array}%
\end{equation}%
Furthermore, their $\rho _{2}$ are found to be identical.

Let us define an operator $\overset{\wedge }{X}$ so that the state $\overset{%
\wedge }{X}|\alpha \rangle $ is related to $|\alpha \rangle $\ by changing
every $k_{i}$\ in $|\alpha \rangle $\ to $-k_{i}+1$, i.e., $\phi
_{k_{i}}(\theta )$ to $\phi _{-k_{i}+1}(\theta )=e^{i\theta }\phi
_{-k_{i}}(\theta )$. \ We further found from the numerical data that
\begin{equation}
\begin{array}{lll}
\Psi _{Z,i}^{(N-L_{o})}=\overset{\wedge }{X}\Psi _{Z,i}^{(L_{o})} &  &
\end{array}%
\end{equation}%
holds exactly. \ In fact, $\overset{\wedge }{X}$ causes a reversion of
rotation of each particle plus a collective excitation. It does not cause
any change in particle correlation, therefore $\rho _{2}$\ remains exactly
unchanged. \ Thus the $L$ large states, including the vortex states $L=N$,
can be known from the $L$ small states.

The underlying physics of this finding is the separability of the
Hamiltonian (it is emphasized that the separability is exact as can be
proved by using mathematical induction). Let $\theta _{coll}=\sum_{i}\theta
_{i}/N,$ which describes a collective rotation. \ Then $H=- \frac{1}{N}\frac{%
\partial ^{2}}{\partial \theta _{coll}^{2}}+H_{int},$ where $H_{int}$
describes the relative (internal) motions and does not depend on $\theta
_{coll}$. Accordingly,\ $E^{(L,Z,i)}=L^{2}/N+E_{int}^{(L,Z,i)}$, the former
is for collective and the latter is for relative (internal) motions. The
eigen-states can be thereby separated as $\Psi _{Z,i}^{(L)}=\frac{1}{\sqrt{%
2\pi }}e^{iL\theta _{coll}}\ \psi _{int}^{(L,Z,i)}$. The feature of the
internal states $\psi _{int}^{(L,Z,i)}$ has been studied in [19]. Where it
was found that, for an arbitrary $L_{o}$
\begin{equation}
\begin{array}{lll}
\psi _{int}^{(N+L_{o},Z,i)}=\psi _{int}^{(L_{o},Z,i)} &  &
\end{array}%
\end{equation}
With these in mind, eq.(10) and (11) can be derived as follows.

From the separability
\begin{equation}
\begin{array}{lll}
\Psi _{Z,i}^{(N-L)}=\frac{1}{\sqrt{2\pi }}e^{i(N-L)\theta _{coll}}\ \psi
_{int}^{(N-L,Z,i)} &  &
\end{array}%
\end{equation}
\ When $\overset{\wedge }{X}$\ acts on a wave function with $L$, from the
definition of $\overset{\wedge }{X}$, $L$ should be changed to $-L$ and an
additional factor $\prod\limits_{j}e^{i\theta _{j}}=e^{iN\ \theta _{coll}}$
should be added, thus
\begin{equation}
\begin{array}{lll}
\overset{\wedge }{X}\Psi _{Z,i}^{(L)}=\frac{1}{\sqrt{2\pi }}e^{i(N-L)\theta
_{coll}}\ \psi _{int}^{(-L,Z,i)} &  &
\end{array}%
\end{equation}
Due to (12), the right hand sides of (13) and (14) are equal, thereby (11)
is proved.

Furthermore, since $\psi _{int}^{(-L,Z,i)}=(\psi _{int}^{(L,Z,i)})^{\ast }$,
the internal energy $%
E_{int}^{(N-L,Z,i)}=E_{int}^{(-L,Z,i)}=E_{int}^{(L,Z,i)} $. \ \ Therefore, $%
E^{(N-L,Z,i)}-E^{(L,Z,i)}=(N-L)^{2}/N-L^{2}/N=N-2L$. This recovers eq.(10),
the energy difference arises purely from the difference in collective
rotation.

If the particles are tightly confined on the ring, rapidly rotating state
with a large $L=JN-L_{o}$ would exist, where $J$ is an integer. \ Their
spectra would remain the same but shift upward by $J(JN-2L_{o})$ from the
spectrum with $L=L_{o},$ while $\Psi _{Z,i}^{(JN-L_{o})}=\overset{\wedge }{
X_{J}}\Psi _{Z,i}^{(L_{o})}$, where $\overset{\wedge }{X_{J}}$ changes each $%
\phi _{k_{i}}$ to $\phi _{-k_{i}+J}$. \ Thus the rapidly rotating states
have the same internal structure as the corresponding lower states but have
a much stronger collective rotation.

When $N$ increases greatly while $V_{o}$ or $\theta _{range}$ decreases
accordingly, the qualitative behaviors remain unchanged. E.g., when $N=10000$
and $V_{o}=0.01$ ($\theta _{range}$ remains unchanged), the spectrum and the
wave functions are found to be nearly the same as the case $N=100$ and $%
V_{o}=1$, except that the spectrum has shifted upward nearly as a whole by
3939. This is again a signal that, for weak interaction and for the ground
states, the mean-field theory is a good approximation.

It is noted that the confinement by a ring is quite different from a
2-dimensional harmonic trap. \ In the latter, the energy of a particle in
the lowest Landau levels is proportional to its angular momentum $k$.
However, for the rings, it is proportional to $k^{2}$. \ Consequently,
higher partial waves are seriously suppressed and the p-wave excitation
becomes dominant. For a harmonic trap it was found in [10,11]\ that d- and
f-wave excitations are more important than the p-wave excitation when $L$\
is small. \ This situation does not appear in our case.

When the zero-range interaction $V_{ij}=g\delta (\theta _{i}-\theta _{j})$
is adopted, The results are nearly the same with those from the
square-barrier interaction if the parameters are related as $g=2V_{o}\theta
_{range}$ (in this choice both interactions have the same diagonal matrix
elements). For an example, a comparison is made in Table III. \ The high
similarity between the two sets of data imply that the above findings are
also valid for zero-range interaction.
\begin{table}[htbp]
\caption{Eigen-energies of the four lowest $L=0$ states for a system with $%
N=100$ and with zero-range interaction $V_{ij}=0.05\protect\delta (\protect%
\theta _{i}-\protect\theta _{j})$ (the unit of energy is $G$\ as before).
The weights of the $j=0$ components of these states are also listed. The
corresponding results from square-barrier interaction with $V_{o}=1$, and $%
\protect\theta _{range}=0.025$ are given in the parentheses.}%
\begin{ruledtabular}
\begin{tabular}{lcc}
 $(L,Z,i)$  &  $E^{(L,Z,i)}$     & $(C_{0}^{(L,Z,i)})^{2}$ \\ \hline
 $(0,I,1)$  &  39.0900 (39.0902) &  0.9370 (0.9371)        \\
 $(0,I,2)$  &  42.2982 (42.2983) &  0.0510 (0.0510)        \\
 $(0,I,3)$  &  45.4733 (45.4733) &  $<$0.02 ($<$0.02)      \\
 $(0,II,1)$ &  47.0076 (47.0074) &  0.8372 (0.8373)
\end{tabular}
\end{ruledtabular}
\end{table}

The numerical results from using zero-range interaction can be compared with
the exact results from solving integral equations by Lieb and Liniger
[16,17]. The variables $\gamma $\ and $e(\gamma )$\ introduced in [16] are
related to those of this paper as $\gamma =g\pi /N$ and $e(\gamma )=4\pi
^{2}E/N^{3}$ (the unit of $E$ is $G$). However, this paper concerns mainly
the case of weak interaction, say, $g\leq 0.05,$or $\gamma \leq 0.00157$
(otherwise, the procedure of diagonalization would not be valid due to the
cutoff of the space). \ Nonetheless, even $\gamma $\ is as large as 0.5 ($%
g=15.9$) the evolution of the ground state energy with $N=100$\ against $%
\gamma $ obtained via diagonalization coincide, in the qualitative sense,
with the exact results quite well . \ This is shown in Fig.8 to be compared
with Fig.3 of [16], where $\gamma $\ is ranged from 0 to 10. \ In Fig.8, the
constraint $\gamma <e(\gamma )$ is recovered. Furthermore, when $\gamma $\
is small, $e(\gamma )$ against $\gamma $\ appears as a straight line.
\begin{figure}[tbph]
\centering
\includegraphics[totalheight=2.0in,trim=30 10 5 10]{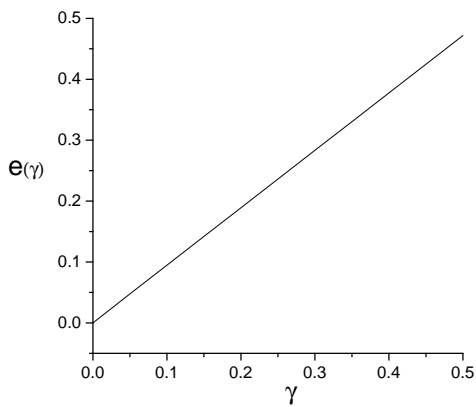}
\caption{$e(\protect\gamma )=4\protect\pi ^{2}E/N^{3}$ against $\protect%
\gamma =g\protect\pi /N$. $N$\ is given at 100 and $E$ is the ground state
energy calculated from the diagonalization in the unit $G$. }
\end{figure}

\bigskip

In summary, a detailed analysis based on the numerical data of $N-$boson
systems on a ring with weak interaction has been made. The main result is
the discovery of the basic pairs, which exist extensively in all the excited
states and dominates the low-lying spectra. \ The fluctuation of basic pairs
provides a common mechanism of oscillation, the low-lying states are thereby
classified into oscillation bands. Each band is characterized by having its
specific additional excitation of a few particles. Since the mechanism of
oscillation is common, the level spacings of different bands are nearly
equal in a spectrum.

To divide the motion into being collective and relative provides a better
understanding to the relation between the higher and lower states. \ The
very high vortex states\ with $L\approx N$\ \ can be understood from the
corresponding low-lying states because they have exactly the same internal
states.

The particle correlation has been intuitively studied. particle densities
are found to be in general non-uniform, bose-clustering originating from the
symmetrization of wave functions is found, which leads to the appearance of
one, two, and three clusters. This phenomenon would become explicit and
might be observed if $N$ is small.

\bigskip

Acknowledgment: \ The support by NSFC under the grants 10574163 and 90306016
is appreciated.

\bigskip

REFERENCES

1, D.M. Stamper-Kurn, M.R. Andrews, A.P. Chikkatur, S. Inouye, H.-J.
Miesner, J. Stenger, and W. Ketterle, \ Phys. Rev. Lett. 80, 2027 (1998)

2, B.P. Anderson and M.A. Kasevich, \ Science 281, 1686 (1998)

3, J.L. Roberts, et al, Phys. Rev. Lett. 81, 5159 (1998)

4, J. Stenger, et al, Phys. Rev. Lett. 82, 2422 (1999).

5, S.L. Cornish et al., Phys. Rev. Lett. 85, 1795 (2000)

6, M. Greiner et al., Nature (London) 415, 39 (2002)

7, B. Paredes et al., Nature (London) 429, 277 (2004)

8, G.T. Kinoshita, T. Wenger, and D.S. Weiss, Science 305, 1125 (2004)

9, N.K. Wilkin, J.M.F. Gunn, and R.A. Smith, Phys. Rev. Lett. 80, 2265 (1998)

10, B. Mottelson, Phys. Rev. Lett. \ 83, 2695 (1999)

11, G.F. Bertsch and T. Papenbrock, 83, 5412 (1999)

12, K. Sakmann, A.I. Streltsov, O.E. Alon, L.S. Cederbaum, Phys. Rev. A 72,
033613 (2005)

13, I. Romanovsky, C. Yannouleas, and U. Landman, Phys. Rev. Lett. 93,
230405 (2004)

14, I. Romanovsky, C. Yannouleas, L.O. Baksmaty, and U. Landman, Phys. Rev.
Lett. 97, 090401 (2006)

15, Yongle Yu, cond-mat/0609711 v1.

16, E.H. Lieb and W.Liniger, Phys, Rev. 130, 1605 (1963)

17, E.H. Lieb, Phys, Rev. 130, 1616 (1963)

18, F. Bacher and S. Goudsmit, Phys. Rev., 46, 948 (1934)

19, C.G. Bao, G.M. Huang, and Y.M. Liu, Phys. Rev. B 72, 195310 (2005)

\end{document}